# One-Dimensional, One-Phase and Two-Phase Eulerian Explicit Shock Tube Simulation Code

M. Giselle Fernández-Godino

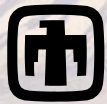

Sandia National Laboratories

# One-Dimensional, One-Phase and Two-Phase Eulerian Explicit Shock Tube Simulation Code

M. Giselle Fernández-Godino
Computer Science Research Institute
Sandia National Laboratories
P.O. Box 5800
Albuquerque, NM 87185-9999
mgferna@sandia.gov

**Abstract**

In this work, a one-dimensional simulation code was developed for both single-phase and two-phase systems. We consider the time-dependent Euler equations for gas and particles. The Euler equations are non-linear hyperbolic conservation laws that govern the dynamics of a compressible material, such as gases or liquids at high pressures, where the effects of body forces, viscous stresses, and heat flux are neglected. The Euler equations were discretized using the finite volume method, and the code was written in MATLAB.

The Sod shock tube problem, a physical analogue of the Riemann problem, is widely used to test the accuracy of computational fluid codes and was also used in this work for that purpose. A discontinuity in pressure was modeled where high and low-pressure regions were separated by a diaphragm. The gas on both sides of the diaphragm is initially at rest, and the density is also discontinuous across the diaphragm. At $t = 0$, the diaphragm breaks. This and other problems were modeled for code verification using exact solutions.

The problem of our interest is a curtain of particles hit by a shock wave due to the significant interest in this phenomenon within the multiphase heterogeneous cylindrical explosion studies by the PSAAP II project. This project was initiated by the US Department of Energy (DOE) National Nuclear Security Administration (NNSA) Office of Advanced Simulation and Computing (ASC), in collaboration with Sandia National Laboratories and the Center for Compressible Multiphase Turbulence at the University of Florida.

Propagation of uncertainties of the selected quantity of interest, maximum density in the particle curtain, was carried out by varying two input variables: initial curtain thickness and initial high density. Uncertainty propagation is often computationally expensive because multiple code evaluations are needed. To address this, a multi-fidelity surrogate model was implemented, combining low and high-fidelity simulations. This model was used to propagate uncertainties with the software DAKOTA, a flexible and extensible interface between analysis codes and iterative systems analysis methods.

# Contents

# List of Figures

# List of Tables

# Chapter 1

# Nomenclature

*x*: Variable in the stream-wise direction.
*t*: Time.
*cell*: Fluid computational domain of a fixed size. Internal cells are fully occupied by the gas, external cells are located outside of the fluid domain and are called ghost cells.
*i*: Subscript that refers to the *i* cell.
$i \pm \frac{1}{2}$: Subscript that refers to the edges of the *i* cell.
*n*: Subscript that refers to the *n* time.
Δ*t*: Time step used in the simulation.
*ρ*: Density.
*u*: Velocity.
*P*: Pressure.
*e*: Internal energy.
*R*: Ideal gas constant.
*γ*: Specific heat ratio.

This page intentionally left blank.

# Chapter 2

# Introduction

## Motivation

The aim of this work is to combine the problem investigated by the Center for Compressible Multiphase Turbulence (CCMT) with a severe accident nuclear reactor simulator that the Computer Science Research Institute at Sandia National Laboratories is developing. CCMT is interested in understanding the instabilities associated with the cylindrical dispersal of 120 μm diameter dry glass particles and 200 μm diameter steel beads [10,11,13,14,]. The problem is being investigated using one, two, and three-dimensional simulations. The purpose of the present work is to develop a simplified one-dimensional simulation code to apply uncertainty quantification methods. The case study includes:

- The development of a one-dimensional simulation code to recreate a curtain of particles being hit by a shock.
- Verification of the simulation code.
- Determination of representative prediction metrics.
- Development of a multi-fidelity surrogate model.
- Uncertainty propagation process.

## The Model

One-dimensional, time-dependent Euler equations with an ideal Equation of State (EOS), using a non-conservative formulation, were implemented. The Euler equations result from neglecting the effects of viscosity, heat conduction, and body forces on a compressible medium. For smooth solutions, all formulations are equivalent. However, for solutions containing discontinuities, such as shock waves, non-conservative formulations give incorrect shock solutions. Nevertheless, non-conservative formulations have some advantages over the conservative counterpart when analyzing the equations or from a numerical point of view [4][8].

The simulation was first modeled as a one-phase system, assuming that the gas was isothermal, and consequently, the energy equation was omitted from the system. Later, this assumption was relaxed; the gas was modeled as a linearized ideal gas, and the energy equation was included. The linearized equation of state was then replaced by the ideal gas equation. Finally, a two-phase model using the Eulerian-Eulerian approach was developed [6].

We are interested in the simulation of a shock hitting a curtain of particles. The simulation can be considered a classical shock tube problem with a thin particle curtain inside [5]. The two-phase approach will be used to model both particles and gas, treating the particles as volumeless. In the Eulerian-Eulerian approach, the different phases are treated mathematically as interpenetrating continua. Conservation equations for each phase are derived to obtain a set of equations with similar structures for all phases. These equations are closed by providing constitutive relations obtained from empirical information. The two phases can exchange momentum and energy.

## Uncertainties

When modeling reality using simulations, there are always errors associated with it. These errors can arise from different sources, including numerical errors (discretization error, round-off error, truncation error), measurement errors, and model form errors. The verification process aims to demonstrate that the model is correctly implemented in simulations and that the input parameters and the model's logical structure are accurately represented. In this work, verification was carried out for different cases.

Measurement errors are usually propagated in simulations. A way to bound these uncertainties is to take samples from the input uncertainty domain and perform simulations for each sample to obtain the interval of output uncertainty. This process can be computationally prohibitive, so surrogate models are often implemented. Sometimes, even the surrogate model of a high-fidelity simulation can be costly. In such cases, multi-fidelity surrogate models are a good option. Multi-fidelity surrogate models are created by combining less accurate but cheaper models with expensive high-fidelity models. In this work, multi-fidelity surrogate models were constructed and used to propagate input uncertainties in simulations.

Model form uncertainty is the error associated with the erroneous selection of the physical model. For example, using a non-conservative formulation, which does not accurately represent shock location and strength, can be treated as a model form uncertainty. In this work, this type of error will not be analyzed.

# Chapter 3

# Single-phase: Governing Equations

*The Eulerian approach was used for modeling the single-phase, and the partial differential equations used are listed below. This includes the derivation of the fundamental equations to the simplified version used.*

## The Equations

The compressible Euler equations are solved using the finite-volume methodology. Neglecting the effects of viscosity and conductivity, we solve the compressible continuity, momentum, and energy equations in the Eulerian framework.

### Conservation of Mass

The conservation of mass for a single-phase is given by:

$$\frac{\partial \rho}{\partial t} + \frac{\partial (\rho u)}{\partial x} = 0, \tag{3.1}$$

where $\rho$ is density, $u$ is velocity, $t$ is time, and $x$ is position.

### Conservation of Momentum

The conservation of momentum for a single-phase is given by

$$\frac{\partial (\rho u)}{\partial t} + \frac{\partial (\rho u^2)}{\partial x} + \frac{\partial P}{\partial x} = 0, \tag{3.2}$$

where $P$ is pressure. Taking the time derivatives of Eq. 3.2 we have,

$$\frac{\partial u}{\partial t} + u \left[ \frac{\partial \rho}{\partial t} + \frac{\partial (\rho u)}{\partial x} \right] + \rho u \frac{\partial u}{\partial x} + \frac{\partial P}{\partial x} = 0. \tag{3.3}$$

Using Eq. 3.1, the bracket expression in 3.3 becomes zero and we have,

$$\frac{\partial u}{\partial t} + u\frac{\partial u}{\partial x} + \frac{1}{\rho}\frac{\partial P}{\partial x} = 0. \tag{3.4}$$

## Conservation of Energy

The conservation of energy for a single-phase is given by

$$\frac{\partial(\rho e)}{\partial t} + \frac{\partial(\rho u e)}{\partial x} + P\frac{\partial u}{\partial x} = 0, \tag{3.5}$$

where $e = C_v\, T$ is the internal energy of the single-phase, where $C_v$ is the specific heat at constant volume and T is the temperature of the single-phase. Taking time derivatives of Eq. 3.5 we obtain

$$e\frac{\partial \rho}{\partial t} + \rho\frac{\partial e}{\partial t} + ue\frac{\partial \rho}{\partial x} + \rho e\frac{\partial u}{\partial x} + \rho u\frac{\partial e}{\partial x} + P\frac{\partial u}{\partial x} = 0, \tag{3.6}$$

Rearranging the expression 3.6 and using 3.1 we have,

$$\frac{\partial e}{\partial t} + u\frac{\partial e}{\partial x} + \frac{P}{\rho}\frac{\partial u}{\partial x}. \tag{3.7}$$

## Equation of State

The ideal gas model was used as the equation of state

$$P = \rho R T = (\gamma - 1)\rho e \tag{3.8}$$

where $\gamma = 1.4$ is the heat capacity ratio of the gas and $R = 8.314\ J/K\cdot mol$ is the gas constant.

# Chapter 4

# Single-phase: Discretization

*The method used to solve the problem is explicit, and the program was written in MATLAB (Appendix A). The finite volume structure of the program is designed for a one-dimensional channel in the axial direction with $n$ number of cells, as shown in Figure 4.1. The first and last cells are ghost cells and act as the boundary conditions. Pressure and density are averaged over the cell volume and are located at the center of the cell, while velocity is located at the faces between the cells. The cells are represented with an index $i$, and the faces with indexes $i + 1/2$ or $i - 1/2$.*

## Initial and Boundary Conditions

The physical phenomenon represented is the propagation of a shock wave. The initial state consists of two tube regions, one with high pressure and the other with low pressure, separated by a diaphragm. The gas on both sides of the diaphragm is initially at rest. The pressure and density are discontinuous across the diaphragm. At $t = 0$, the diaphragm is broken. Two types of singularities then propagate through the gas: contact discontinuities, where the pressure and velocity are continuous, but density and energy are discontinuous, and shock waves, where all quantities—pressure, velocity, density, and energy—are discontinuous across the shock front.

Figure 4.1 shows the mesh configuration; pressure, energy, and density are calculated at the cell center, while velocity is calculated at the cell edges. Ghost cells were defined at the extremes of the mesh; these cells are needed to compute the calculations in the first and last real cells.

Initial boundary conditions must satisfy the equation of state given in equation 3.8, which provides a relationship between density ($\rho$), pressure (P), and energy ($e$). The velocity ($u$) is initially set to zero at the cell edges and in ghost cells for all times. The one-dimensional equations are then evaluated at a position index $i$ and a certain time $n$ to solve for the next time value $n + 1$.

## Discrete Equations

Finite differences have been used to discretize the analytic equations using control volume integration. The resulting equations are listed below.

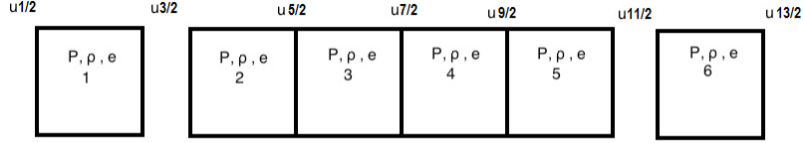

**Figure 4.1.** Staggered mesh showing the location of pressure, density, internal energy and velocity conditions for six cells.

## Conservation of Mass

$$\rho_i^n = \rho_i^{n-1} - \frac{\Delta t}{\Delta x}\left[\rho_{out} u_{i+1}^{n-1} - \rho_{in} u_i^{n-1}\right] \tag{4.1}$$

where

$$\rho_{out} = \begin{cases} \rho_i^{n-1} & \text{if } u_{i+1}^{n-1} > 0 \\ \rho_{i+1}^{n-1} & \text{otherwise} \end{cases} \tag{4.2}$$

and

$$\rho_{in} = \begin{cases} \rho_{i-1}^{n-1} & \text{if } u_i^{n-1} > 0 \\ \rho_{i+1}^{n-1} & \text{otherwise} \end{cases}. \tag{4.3}$$

## Conservation of Momentum

$$u_{i+\frac{1}{2}}^n = u_{i+\frac{1}{2}}^{n-1} - \frac{\Delta t}{\Delta x}\left[u_{i+\frac{1}{2}}^{n-1} \delta u_{i+\frac{1}{2}}^{n-1} + \frac{1}{\rho_{i-\frac{1}{2}}^{n-1}}\left(P_{i+\frac{1}{2}}^n - P_{i-\frac{1}{2}}^n\right)\right] \tag{4.4}$$

$$\delta u_{i+\frac{1}{2}}^{n-1} = \begin{cases} u_{i+\frac{1}{2}}^{n-1} - u_{i-\frac{1}{2}}^{n-1} & \text{if } u_{i+\frac{1}{2}}^{n-1} > 0 \\ u_{i+\frac{3}{2}}^{n-1} - u_{i+\frac{1}{2}}^{n-1} & \text{otherwise} \end{cases} \tag{4.5}$$

and

$$\bar{\rho}_{i-\frac{1}{2}}^{n-1} = \frac{1}{2}(\rho_i^{n-1} + \rho_{i-1}^{n-1}) \tag{4.6}$$

## Conservation of Energy

$$e_i^n = e_i^{n-1} - \frac{\Delta t}{\Delta x}\left[\bar{u}\delta e + \frac{P_i^{n-1}}{\bar{u}}(u_{i+1}^{n-1} - u_i^{n-1})\right] \tag{4.7}$$

$$\delta e = \begin{cases} e_i^{n-1} - e_{i-1}^{n-1} & \text{if } u_i^{n-1} > 0 \\ e_{i+1}^{n-1} - e_i^{n-1} & \text{otherwise} \end{cases} \tag{4.8}$$

and

$$\bar{u} = \frac{1}{2}(u_{i+1}^{n-1} + u_i^{n-1}) \tag{4.9}$$

## Equation of State

The ideal gas model was used as the equation of state

$$P_i^n = (\gamma - 1)\rho_i^n e_i^n. \tag{4.10}$$

This page intentionally left blank.

# Chapter 5

# Single-phase: Code Verification

*Code verification uses quantitative comparisons of numerical and exact solutions to identify errors in algorithm implementation and algorithmic weaknesses. The primary focus is on the order of convergence of the error as a function of discretization parameters.*

## The Riemann Problem

A Riemann problem consists of an initial value problem composed of a conservation equation together with piecewise constant data having a single discontinuity. The Riemann problem is very useful for understanding equations like the Euler conservation equations because all properties, such as shocks and rarefaction waves, appear as characteristics in the solution. It also provides an exact solution to complex nonlinear equations, such as the Euler equations. In numerical analysis, Riemann problems naturally appear in finite volume methods for the solution of conservation law equations due to the discreteness of the grid. For this reason, it is widely used in computational fluid dynamics and magnetohydrodynamics simulations. In these fields, Riemann problems are solved using Riemann solvers.

### Exact Solution

The Riemann problem is also known as the shock tube problem. A shock tube is created by setting the initial mass flow rate and velocity to zero with no gravity [9, 12]. The boundary conditions at the inlet and outlet are also zero (closed system). One half of the domain is defined as a high-pressure region, while the other half is the low-pressure region, separated by a diaphragm. At $t = 0$, the diaphragm disappears. When the diaphragm disappears, a compression wave moves to the right, and a rarefaction wave moves to the left. The domain is then split into four different regions. The region to the left of the rarefaction wave has the same properties as the initial right region. For a perfect caloric gas, the following equations are provided given the initial conditions for states 1 and 4 (see Fig. 5.1).

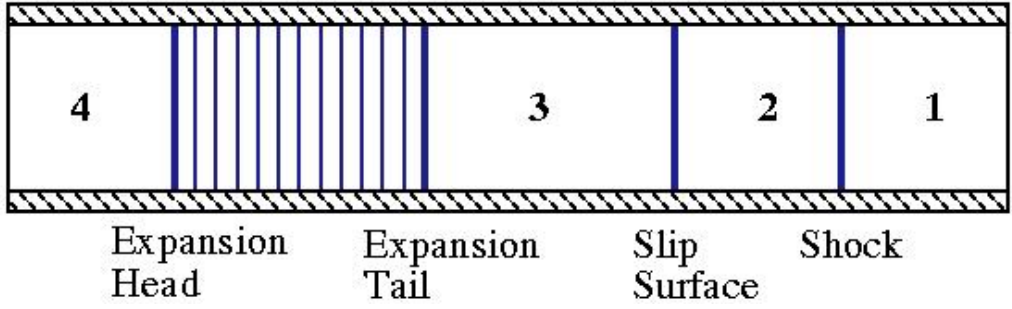

**Figure 5.1.** Shock tube shortly after the diaphragm has burst.

## Verification Process

It is necessary to keep in mind that the discretized equations in the present simulations were constructed using primitive variables ($u$, $e$) instead of the conserved variables ($\rho u$, $\rho(\frac{1}{2} u^2 + e)$). For smooth solutions, all formulations are equivalent. However, for solutions containing shock waves, non-conservative formulations give incorrect shock solutions [7]. The method used has an order of convergence of one, but because the test case has a discontinuity (the shock wave), a smaller order of convergence is expected. Moreover, because the test case is not smooth, the exact solution will never be reached.

The verification was performed by setting the initial gas velocity to zero. The gas pressure gradient was set to 0.5 ($P_{g_0 L} = 1$ and $P_{g_0 R} = 0.5$). In order to calculate the error in each variable, we first determine the corresponding value using the exact solution, and then we compute the results using the one-dimensional code. Once we have these two values for all the cells and interfaces, we calculate the error as the sum of the absolute value of the difference between the exact solution and the one-dimensional code, divided by the number of cells/interfaces. This is expressed in Eq. 5.1 for cell quantities and in Eq. 5.2 for edge quantities.

$$error_i = \frac{1}{n} \sum_{i=1}^{n} |y_{exact_i} - y_{sim_i}|, \qquad (5.1)$$

$$error_{i+\frac{1}{2}} = \frac{1}{n} \sum_{i=\frac{1}{2}}^{n+\frac{1}{2}} |y_{exact_{i+\frac{1}{2}}} - y_{sim_{i+\frac{1}{2}}}|, \qquad (5.2)$$

where $y_{exact_i}$ is the result of the exact solution at the point $i$, $y_{sim_i}$ is the simulation code result at point $i$ and $n$ is the number of cells considered. Similarly, this applies to the edge quantities.

Here, we used the L1 norm because for a discontinuous phenomenon, the L1 norm is the unique norm that converges. The method used has a convergence coefficient of one, but when the test cases have discontinuities, such as a shock wave, a smaller order of convergence is expected.

## Verification for the Sod Test

Convergence to the exact solution is not expected when discontinuities are present, as in the Sod test. To test the simulation code, which includes the ideal gas equation of state, we consider the Sod problem. This is an example of a Riemann problem, where the initial conditions are defined by two constant states separated by an interface. At t=0t = 0t=0, the interface is broken, leading to a self-similar set of waves that propagate through the domain. Riemann problems for the single-phase Euler equations and the ideal gas EOS have exact solutions that can be computed with arbitrarily small error by iteratively solving a scalar nonlinear equation. For the Sod problem, we set the initial conditions to: $\rho_L$ = 1, $\rho_R$ = 0.125, $P_L$ = 1, $P_R$ = 0.1 and $u_L$ = $u_R$ = 0. The test can be found on page 129 of reference [7] and the procedure for computing the exact solution is in the same reference. The results are listed below.

The shock tube length was set to $x = 1$ $m$, with the diaphragm located at $x = 0.5$ $m$. Thetime step is variable during the simulation and is set as

$$\Delta t = \frac{\text{CFL } \Delta x}{\text{Max}(a+u)} \quad (5.3)$$

where CFL denotes the Courant-Friedrichs-Lewy number and is set to 0.9. Figure 5.2 shows the comparison between the exact solution inspired by reference [7], and the results of the single-phase code with the CFL number fixed at 0.9 and varying $\Delta x$ and $\Delta t$. As expected, there is a difference between the location of the shock in the exact solution and in the single-phase simulation due to the non-conservative formulation used.

Figure 5.2 shows the results for 200 cells, with an end time of 0.25 s, which is the product of the number of time steps and $\Delta t$. Figure 5.3 shows how the errors in density, pressure, velocity, and internal energy are reduced as $\Delta x$ is reduced. The optimal convergence is not reached because we discretized the primitive variables $u$ and $e$ instead of the conserved variables $\rho u$ and $\rho e$, resulting in a lack of discrete conservation. As discussed earlier, a convergence ratio smaller than one was expected. The results showed a value around 0.5.

The single-phase code can be improved by using a conservative formulation of the discretized equations. However, because we are interested in including particles as a second phase, we consider that the error due to the model will be negligible in comparison with other error sources. Therefore, the conservative formulation will not be covered in this report.

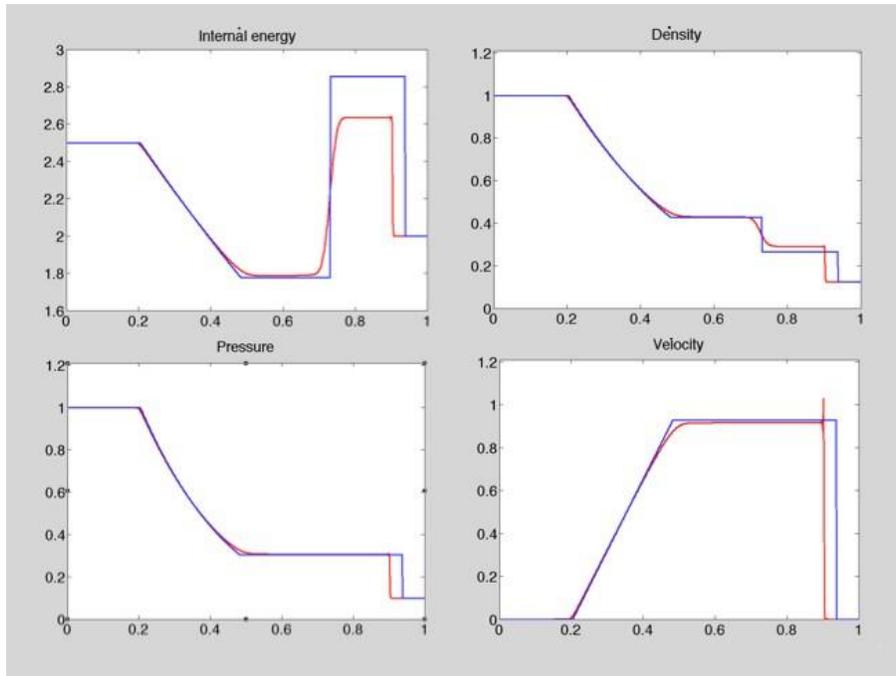

**Figure 5.2.** Comparison between the exact solution (blue) and the single-phase code (red) for internal energy, density, pressure, and velocity. The model used is not accurate enough to determine the shock position precisely.

## Smaller Pressure Gradient Sod Problem

A second verification process was performed using the same initial and boundary conditions as in the Sod test, but with a different pressure gradient. The case considered has a smaller pressure gradient, resulting in a smaller discontinuity and a better convergence rate is expected. The alternative case was set using a density gradient of $\rho_L = 1$, $\rho_R = 0.125$, and pressure gradient of $P_L = 1$, $P_R = 0.5$ with $u_L = u_R = 0$. In Figure 5.4 better prediction of the shock location is observed.

## CFL number fixed

Figure 5.5 shows the comparison between the exact solution and the results of the single-phase code with a fixed CFL number of 0.9 and varying $\Delta x$ and $\Delta t$. The relationship is linear, and the convergence coefficient is around 0.8.

| $n_{cells}$ | $\Delta x$ | $error_\rho$ | $error_e$ | $error_P$ | $error_u$ |
|---|---|---|---|---|---|
| 20 | 0.05 | 0.0683 | 0.7535 | 0.0918 | 0.2109 |
| 40 | 0.025 | 0.0402 | 0.4568 | 0.0561 | 0.1252 |
| 80 | 0.0125 | 0.0238 | 0.2922 | 0.0288 | 0.0644 |
| 160 | 0.00625 | 0.0157 | 0.1864 | 0.0157 | 0.0339 |
| 320 | 0.003125 | 0.0101 | 0.1325 | 0.0095 | 0.0203 |
| 640 | 0.0015625 | 0.0072 | 0.0856 | 0.0061 | 0.0136 |

**Table 5.1.** Errors calculated for density, internal energy, pressure, and velocity. The results were obtained with a CFL number fixed at 0.9 and varying $\Delta t$ and $\Delta x$.

| $n_{cells}$ | $p_\rho$ | $p_e$ | $p_P$ | $p_u$ |
|---|---|---|---|---|
| 20-40 | 0.764690077 | 0.722044861 | 0.710493383 | 0.752324533 |
| 40-80 | 0.756233928 | 0.644606473 | 0.961931959 | 0.959101969 |
| 80-160 | 0.600197014 | 0.648554318 | 0.875304253 | 0.925775415 |
| 160-320 | 0.636409266 | 0.4924095 | 0.724765141 | 0.739805546 |
| 320-640 | 0.488286481 | 0.630309658 | 0.639118271 | 0.577873076 |

**Table 5.2.** Convergence exponent, p, for density, internal energy, pressure and velocity. The results were obtained with a CFL number fixed at 0.9 and varying $\Delta t$ and $\Delta x$.

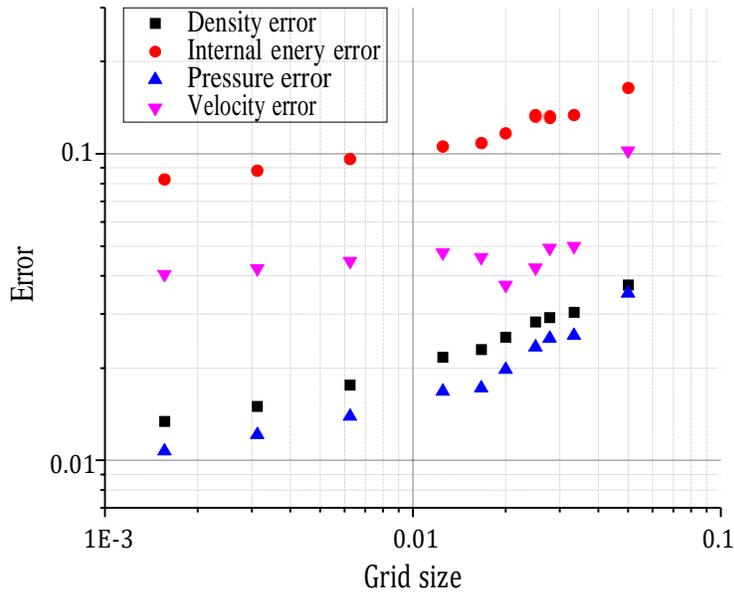

**Figure 5.3.** CFL was maintained constant at a value of 0.9 varying Δ*t* and Δ*x*.

## Time Step Fixed

Figure 5.6 shows the comparison between the exact solution and the results of the single-phase code with Δt fixed at 1x10−4 and varying CFL number and Δx. The relationship is linear, and the convergence coefficient is around 0.8.

# Verification for the Sine Wave

For smooth solutions, like the sine wave, the one-dimensional single-phase simulation code is expected to converge to the exact solution at a rate of one (first-order convergence). The exact solution was obtained from [7]. The sine wave was modeled with periodic boundary conditions, and the simulation code was verified. Figure 5.7 shows the exact solution against the one-dimensional single-phase simulation code for 0.25 seconds using 1280 cells. The solutions match, and no apparent differences exist.

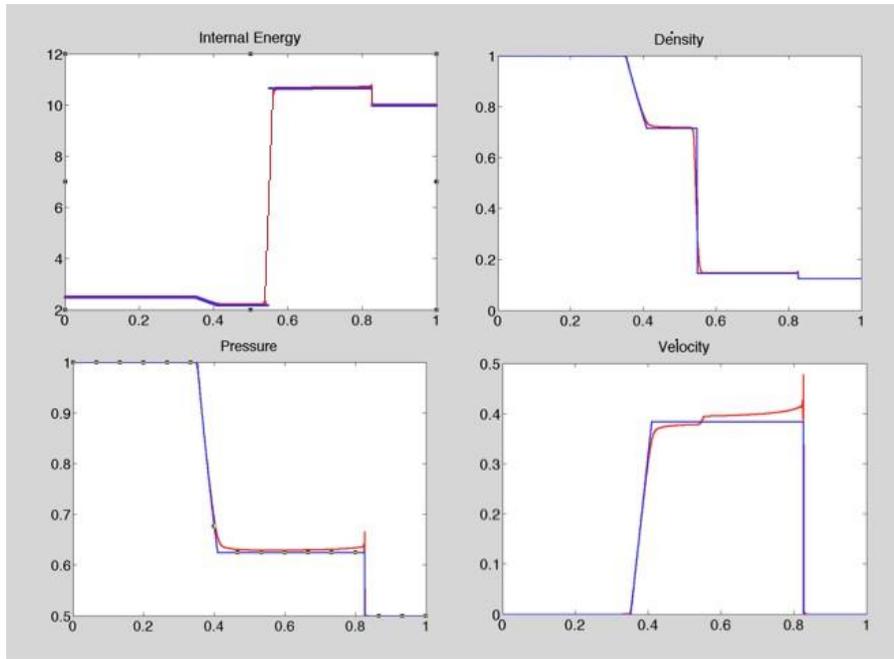

**Figure 5.4.** Comparison between the exact solution (blue) and the single-phase code (red) for internal energy, density, pressure and velocity. The comparison between the exact solution and the simulations shows a better determination of the shock location.

## CFL number fixed

Figure 5.8 shows the comparison between the exact solution and the results of the single-phase code with the CFL number fixed at 0.9 and varying Δ$t$ and Δ$x$. The relationship is linear and the order of convergence, p, is around 1 as expected. (see Table 5.4).

## Time Step Fixed

Figure 5.9 shows a comparison between the exact solution and the results of the single-phase code with Δ$t$ fixed at 1$x$10$^{-4}$ while varying CFL number and Δ$x$. The relationship is linear and the exponent of convergence, p, is around 1 as it was expected (see Table 5.6).

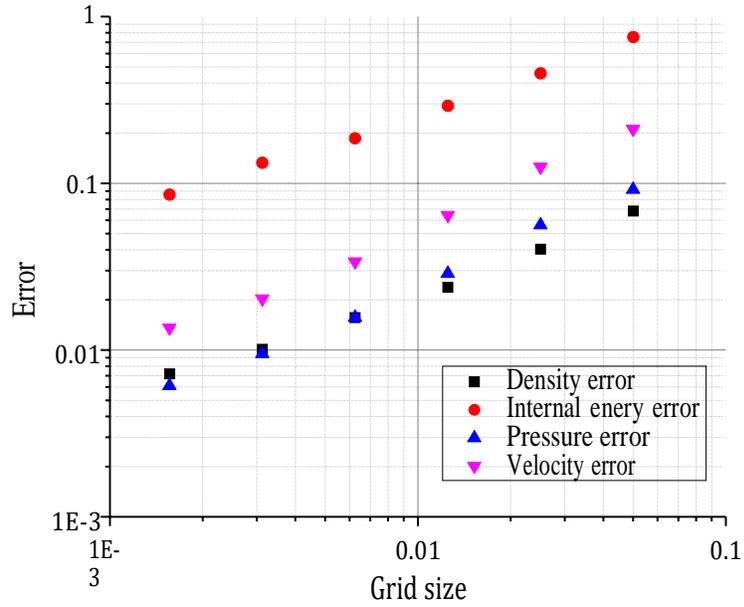

**Figure 5.5.** CFL was maintained constant at a value of 0.9 while Δ*t* and Δ*x* were varied.

| $n_{cells}$ | $error_\rho$ | $error_P$ | $error_e$ | $error_u$ |
|---|---|---|---|---|
| 10 | 0.108195 | 0.079788 | 0.430568 | 0.009215 |
| 20 | 0.050850 | 0.044298 | 0.215005 | 0.007114 |
| 40 | 0.026128 | 0.023805 | 0.111577 | 0.004486 |
| 80 | 0.013998 | 0.012338 | 0.058255 | 0.002540 |
| 160 | 0.006364 | 0.006229 | 0.027414 | 0.001346 |
| 320 | 0.003292 | 0.003142 | 0.013988 | 0.000696 |
| 640 | 0.001520 | 0.001575 | 0.006653 | 0.000353 |
| 1280 | 0.000845 | 0.000790 | 0.003549 | 0.000178 |

**Table 5.3.** Errors calculated for density, pressure, internal energy, and velocity. Results obtained with CFL number fixed at 0.9 while varying Δ*t* and Δ*x*.

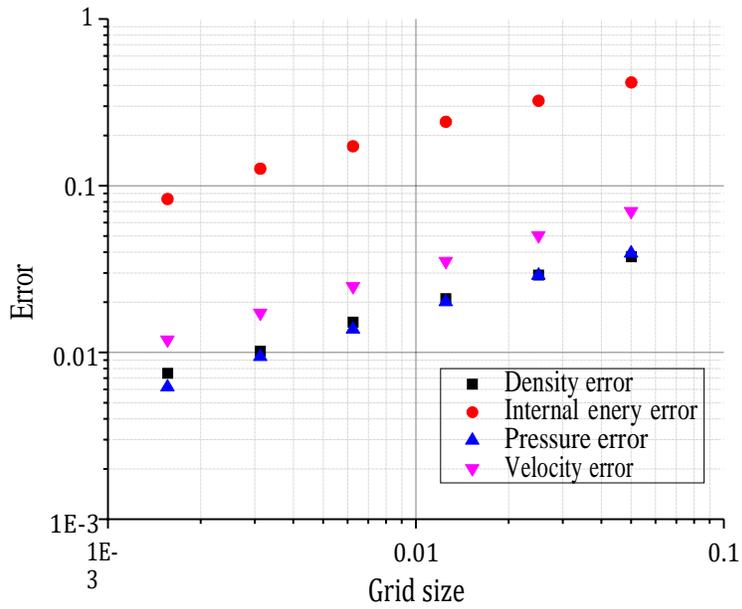

**Figure 5.6.** $\Delta t$ was maintained constant at a value of $1x10^{-4}$ while CFL and $\Delta x$ were varied.

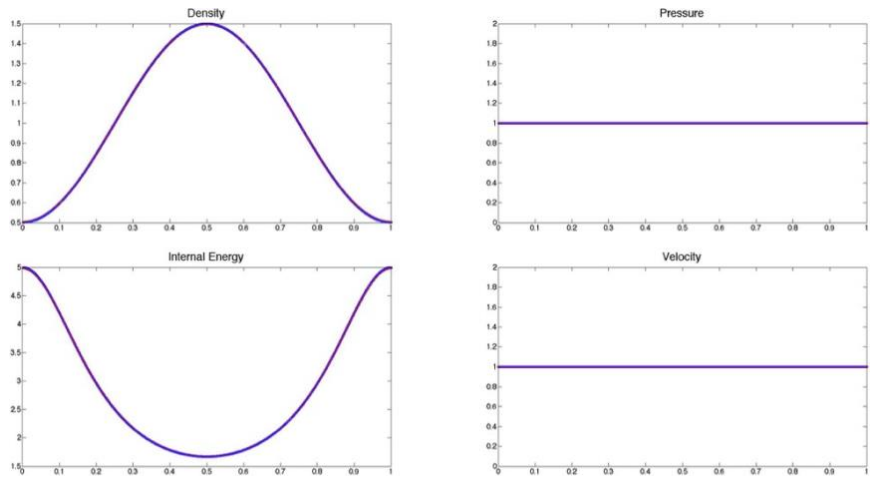

**Figure 5.7.** Qualitative comparison between the exact solution (blue) and the single-phase code (red) for internal energy, density, pressure and velocity.

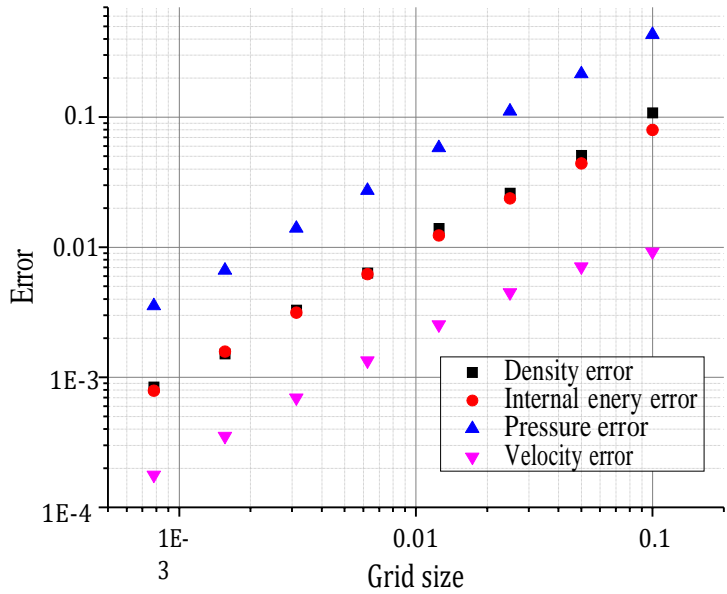

**Figure 5.8.** CFL number was maintained constant at a value of 0.9 while Δ*t* and Δ*x* were varied.

| $n_{cells}$ | $p_\rho$ | $p_P$ | $p_e$ | $p_u$ |
|---|---|---|---|---|
| 10-20 | 1.089317 | 0.848912 | 1.001870 | 0.373298 |
| 20-40 | 0.960667 | 0.895980 | 0.946331 | 0.665321 |
| 40-80 | 0.900393 | 0.948196 | 0.937596 | 0.820449 |
| 80-160 | 1.137227 | 0.985924 | 1.087439 | 0.916244 |
| 160-320 | 0.950916 | 0.987303 | 0.970701 | 0.952152 |
| 320-640 | 1.114886 | 0.996649 | 1.072116 | 0.978714 |
| 640-1280 | 0.847002 | 0.995334 | 0.906689 | 0.986155 |

**Table 5.4.** Convergence exponent, p, for density, pressure, internal energy, and velocity. Results obtained with a CFL number fixed at 0.9 while varying Δ*t* and Δ*x*.

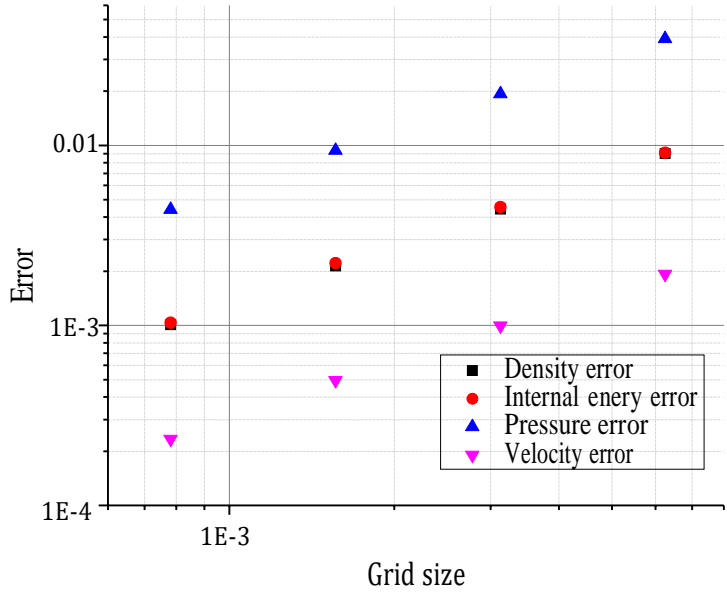

**Figure 5.9.** $\Delta t$ was maintained constant at a value of $1x10^{-4}$ while CFL number and $\Delta x$ were varied.

| $n_{cells}$ | $error_\rho$ | $error_P$ | $error_e$ | $error_u$ |
|---|---|---|---|---|
| 160 | 0.009023 | 0.009115 | 0.039108 | 0.001934 |
| 320 | 0.004437 | 0.004553 | 0.019331 | 0.000999 |
| 640 | 0.002148 | 0.002218 | 0.009379 | 0.000495 |
| 1280 | 0.001011 | 0.001037 | 0.004402 | 0.000234 |

**Table 5.5.** Errors calculated for density, internal energy, pressure and velocity. Results obtained with $\Delta t$ fixed at $1x10^{-4}$ while varying CFL number and $\Delta x$.

| $n_{cells}$ | $p_\rho$ | $p_P$ | $p_e$ | $p_u$ |
|---|---|---|---|---|
| 160-320 | 1.024155 | 1.001631 | 1.016545 | 0.952844 |
| 320-640 | 1.046394 | 1.037416 | 1.043402 | 1.012432 |
| 640-1280 | 1.087737 | 1.096637 | 1.091401 | 1.083795 |

**Table 5.6.** Convergence exponent, p, for density, internal energy pressure and velocity. Results obtained with $\Delta t$ fixed at $1x10^{-4}$ while varying CFL number and $\Delta x$.

This page intentionally left blank.

# Chapter 6

# Two-phases: Partial Differential Equations

*The extension to two phases was done using the Eulerian-Eulerian approach, i.e., the Eulerian approach for both gas and particles. The partial differential equations used are listed below. This includes the derivation of the fundamental equations to the simplified version used, following the same structure as in section 3.*

## Eulerian-Eulerian approach

The fluid phase is always modeled using the Eulerian approach, but generally, there are two methods of modeling particle transport in CFD simulations: the Eulerian method and the Lagrangian method. The Eulerian method treats the particle phase as a continuum and develops its conservation equations on a control volume basis, in a similar form to those for the fluid phase. The Lagrangian method considers particles as a discrete phase and tracks the pathway of each individual particle. By studying the statistics of particle trajectories, the Lagrangian method can also calculate particle concentration and other phase data. When the volume fraction of particles is not too large, the Eulerian-Eulerian approach is very accurate and less computationally expensive than the Eulerian-Lagrangian approach [1].

## The Equations

The conservation of mass, energy, and momentum, along with the equation of state for the two phases (gas g and particles p) are included. The volume of the particles will not be considered.

## Conservation of mass

$$\frac{\partial \rho_g}{\partial t} + \frac{\partial (\rho_g u_g)}{\partial x} = 0, \qquad (6.1)$$

$$\frac{\partial \rho_p}{\partial t} + \frac{\partial (\rho_p u_p)}{\partial x} = 0. \tag{6.1}$$

## Conservation of momentum

$$\frac{\partial (\rho_g u_g)}{\partial t} + \frac{\partial (\rho_g u_g^{\,2})}{\partial x} + \frac{\partial P_g}{\partial x} + F|u_g - u_p|(u_g - u_p) = 0, \tag{6.3}$$

$$\frac{\partial (\rho_p u_p)}{\partial t} + \frac{\partial (\rho_p u_p^{\,2})}{\partial x} + \frac{\partial P_p}{\partial x} - F|u_g - u_p|(u_g - u_p) = 0, \tag{6.4}$$

and taking the time derivatives of 6.3 and 6.4 we have,

$$\rho_g \frac{\partial u_g}{\partial t} + u_g \left[ \frac{\partial \rho_g}{\partial t} + \frac{\partial (\rho_g u_g)}{\partial x} \right] + \rho_g u_g \frac{\partial u_g}{\partial x} + \frac{\partial P_g}{\partial x} + \frac{d}{dt} F|u_g - u_p|(u_g - u_p) = 0, \tag{6.5}$$

$$\rho_p \frac{\partial u_p}{\partial t} + u_p \left[ \frac{\partial \rho_p}{\partial t} + \frac{\partial (\rho_p u_p)}{\partial x} \right] + \rho_p u_p \frac{\partial u_p}{\partial x} + \frac{\partial P_p}{\partial x} - \frac{d}{dt} F|u_g - u_p|(u_g - u_p) = 0, \tag{6.6}$$

Using Eq. 6.1 and Eq. 6.2, the bracket expressions in 6.5 and 6.6 become zero and we have

$$\frac{\partial u_g}{\partial t} + u_g \frac{\partial u_g}{\partial x} + \frac{1}{\rho_g} \frac{\partial P_g}{\partial x} + \frac{d}{dt} F|u_g - u_p|(u_g - u_p) = 0, \tag{6.7}$$

$$\frac{\partial u_p}{\partial t} + u_p \frac{\partial u_p}{\partial x} + \frac{1}{\rho_p} \frac{\partial P_p}{\partial x} + \frac{d}{dt} F|u_g - u_p|(u_g - u_p) = 0. \tag{6.8}$$

## Conservation of energy

$$\frac{\partial (\rho_g e_g)}{\partial t} + \frac{\partial (\rho_g u_g e_g)}{\partial x} + P_g \frac{\partial u_g}{\partial x} + H(e_g - e_p) = 0, \tag{6.9}$$

$$\frac{\partial (\rho_p e_p)}{\partial t} + \frac{\partial (\rho_p u_p e_p)}{\partial x} + P_p \frac{\partial u_p}{\partial x} - H(e_g - e_p) = 0, \tag{6.10}$$

taking time derivatives, equation 6.9 and 6.10 become

$$e_g \frac{\partial \rho_g}{\partial t} + \rho_g \frac{\partial e_g}{\partial t} + u_g e_g \frac{\partial \rho_g}{\partial x} + \rho_g e_g \frac{\partial u_g}{\partial x} + \rho_g u_g \frac{\partial e_g}{\partial x} + P_g \frac{\partial u_g}{\partial x} + \frac{d}{dt} H(e_g - e_p) = 0, \quad (6.11)$$

$$e_p \frac{\partial \rho_p}{\partial t} + \rho_p \frac{\partial e_p}{\partial t} + u_p e_p \frac{\partial \rho_p}{\partial x} + \rho_p e_p \frac{\partial u_p}{\partial x} + \rho_p u_p \frac{\partial e_p}{\partial x} + P_p \frac{\partial u_p}{\partial x} - \frac{d}{dt} H(e_g - e_p) = 0, \quad (6.12)$$

rearranging the expressions 6.11 and 6.12, and using 6.1 and 6.2 we obtain,

$$\frac{\partial e_g}{\partial t} + u_g \frac{\partial e_g}{\partial x} + \frac{P_g}{\rho_g} \frac{\partial u_g}{\partial x} + \frac{d}{dt} H(e_g - e_p) = 0, \qquad (6.13)$$

$$\frac{\partial e_p}{\partial t} + u_p \frac{\partial e_p}{\partial x} + \frac{P_p}{\rho_p} \frac{\partial u_p}{\partial x} + \frac{d}{dt} H(e_g - e_p) = 0. \qquad (6.14)$$

## Equation of state

The ideal gas model was used as equation of state

*Ideal gas model:*

$$P_g = \rho_g R_g T = (\gamma_g - 1)\rho_g e_g \qquad (6.15)$$

where $\gamma = 1.4$ is the heat capacity ratio of the gas and $R_g = 8.314$ $J/K \cdot mol$ is the gas constant.

$$P_p = \rho_p R_p T = (\gamma_p - 1)\rho_p e_p \qquad (6.16)$$

where $\gamma = 1.4$ is the heat capacity for particles and $R_p = 8.314$ $J/K$ $mol$ is the particle material constant.

This page intentionally left blank.

# Chapter 7

# Two-phase: Discretization

*The discretized equations derived from the finite volume method are listed below. The structure of this section follows that used in section 4, including the equations for the second phase.*

## Initial and Boundary Conditions

The first and last cells, called ghost cells, are set with constant values for density ($\rho_g$), pressure ($P_g$), internal energy ($e_g$) and velocity ($u_g$) for all times. Initial boundary conditions must satisfy the equations of state given in equations 6.15 and 6.16, which provide a relationship between $\rho_g$, $P_g$ and $e_g$. In the chosen formulation particles will be treated as gas and they also must satisfy the equation of state used for gas. The velocity $u_g$ and $u_p$ were set to zero at all edges at the initial time and were also set to zero at the ghost edges (i.e., the first two edges and the last two edges) for all times. Particles have no initial density gradient or pressure gradient. The gas will have an initial pressure gradient and density gradient ($\rho_{g_L} = 1$, $\rho_{g_R} = 0.125$, $P_{g_L} = 1$, $P_{g_R} = 0.5$ and $u_{g_L} = u_{g_R} = 0$). In this preliminary approach F (friction coefficient) and H (heat transfer coefficient) were set to zero, and these results were verified in the next section.

If the friction coefficient (F) and the heat transfer coefficient (H) are set to zero, we should obtain two independent solutions: one for gas and the other for particles. The verification was performed by setting the initial velocity to zero for both gas and particles. The gas pressure gradient was set to 0.5 ($P_{g_{0L}} = 1$ and $P_{g_{0R}} = 0.5$) and the particles' pressure gradient was set to zero ($P_{p_{0L}} = 0.5$ and $P_{p_{0R}} = 0.5$).

To calculate the error in each variable we first determine the correspondent value using thIn order to calculate the error in each variable, we first determine the corresponding value using the exact solution and then compute the results using the two-phase simulation code. Once we have these two values for all the cells/edges, we proceed to calculate the error as the sum of the absolute value of the difference between the exact solution and the two-phase simulation code, divided by the number of cells/edges.

# Staggered Mesh

Figure 7.1 shows the discretized mesh used to model the problem.

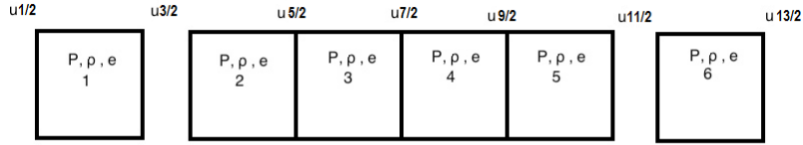

**Figure 7.1.** Staggered mesh showing the location of pressure, density, internal energy and velocity conditions for 6 cells.

# Discrete Equations

Finite differences have been used to discretize the analytic equations using control volume integration. The resulting equations are listed below.

## Conservation of Mass

$$\rho_{g_i}^n = \rho_{g_i}^{n-1} - \frac{\Delta t}{\Delta x}\left[\rho_{g_{out}} u_{i+1}^{n-1} - \rho_{g_{in}} u_{g_i}^{n-1}\right] \tag{7.1}$$

where

$$\rho_{g_{out}} = \begin{cases} \rho_{g_i}^{n-1} & \text{if } u_{g_{i+1}}^{n-1} > 0 \\ \rho_{g_{i+1}}^{n-1} & \text{otherwise} \end{cases} \tag{7.2}$$

and

$$\rho_{g_{in}} = \begin{cases} \rho_{g_{i-1}}^{n-1} & \text{if } u_{g_i}^{n-1} > 0 \\ \rho_{g_{i+1}}^{n-1} & \text{otherwise} \end{cases}. \tag{7.3}$$

*For particles:*

$$\rho_{p_i}^n = \rho_{p_i}^{n-1} - \frac{\Delta t}{\Delta x}\left[\rho_{p_{out}} u_{i+1}^{n-1} - \rho_{p_{in}} u_{p_i}^{n-1}\right] \qquad (7.4)$$

where

$$\rho_{p_{out}} = \begin{cases} \rho_{p_i}^{n-1} & \text{if } u_{p_{i+1}}^{n-1} > 0 \\ \rho_{p_{i+1}}^{n-1} & \text{otherwise} \end{cases} \qquad (7.5)$$

and

$$\rho_{p_{in}} = \begin{cases} \rho_{p_{i-1}}^{n-1} & \text{if } u_{p_i}^{n-1} > 0 \\ \rho_{p_{i+1}}^{n-1} & \text{otherwise} \end{cases}. \qquad (7.6)$$

## Conservation of Momentum

*For gas:*

$$u_{g_i}^n = u_{g_i}^{n-1} - \frac{\Delta t}{\Delta x}\left[u_{g_i}^{n-1}\delta u_{g_i}^{n-1} + \frac{1}{\bar{\rho}_{g_i}^{n-1}}\left(P_{g_i}^n - P_{g_i}^{n-1}\right)\right] - \frac{F\,\Delta t}{\bar{\rho}_{g_i}^{n-1}}\left|u_{g_i}^{n-1} - u_{p_i}^{n-1}\right|\left(u_{g_i}^{n-1} - u_{p_i}^{n-1}\right) \qquad (7.7)$$

(7.7)

$$\delta u_{g_i}^{n-1} = \begin{cases} u_{g_i}^{n-1} - u_{g_{i-1}}^{n-1} & \text{if } u_{g_i}^{n-1} > 0 \\ u_{g_{i+1}}^{n-1} - u_{g_i}^{n-1} & \text{otherwise} \end{cases} \qquad (7.8)$$

and

$$\bar{\rho}_{g_i}^{n-1} = \frac{1}{2}\left(\rho_{g_i}^{n-1} + \rho_{g_{i-1}}^{n-1}\right) \qquad (7.9)$$

*For particles:*

$$u_{p_i}^n = u_{p_i}^{n-1} - \frac{\Delta t}{\Delta x}\left[u_{p_i}^{n-1}\delta u_{p_i}^{n-1} + \frac{1}{\bar{\rho}_{p_i}^{n-1}}\left(P_{p_i}^n - P_{p_i}^{n-1}\right)\right] - \frac{F\,\Delta t}{\bar{\rho}_{p_i}^{n-1}}\left|u_{p_i}^{n-1} - u_{g_i}^{n-1}\right|\left(u_{g_i}^{n-1} - u_{p_i}^{n-1}\right) \qquad (7.10)$$

$$\delta u_{p_i}^{n-1} = \begin{cases} u_{p_i}^{n-1} - u_{p_{i-1}}^{n-1} & \text{if } u_{p_i}^{n-1} > 0 \\ u_{p_{i+1}}^{n-1} - u_{p_i}^{n-1} & \text{otherwise} \end{cases} \qquad (7.11)$$

and

$$\bar{\rho}_{p_i}^{n-1} = \tfrac{1}{2}\left(\rho_{p_i}^{n-1} + \rho_{p_{i-1}}^{n-1}\right) \tag{7.12}$$

**Conservation of Energy**

*For gas:*

$$e_{g_i}^n = e_{g_i}^{n-1} - \tfrac{\Delta t}{\Delta x}\left[\bar{u}_g \delta e_g + \tfrac{P_{g_i}^{n-1}}{\bar{u}_g}\left(u_{g_{i+1}}^{n-1} - u_{g_i}^{n-1}\right)\right] + H\Delta t\left(e_{g_i}^{n-1} - e_{p_i}^{n-1}\right) \tag{7.13}$$

where

$$\bar{u}_{g_i}^n = \tfrac{1}{2}\left(u_{g_i}^{n-1} + u_{g_i}^{n-1}\right) \tag{7.14}$$

and

$$\delta e_{g_i}^n = \bar{u}_{g_i}^n \delta \text{energy}_{g_i}^n + \tfrac{P_{g_i}^{n-1}}{\rho_{g_i}^{n-1}}\left(u_{g_{i+1}}^{n-1} + u_{g_{i+1}}^{n-1}\right) \tag{7.15}$$

$$\delta \text{energy}_{g_i}^n = \begin{cases} e_{g_i}^{n-1} - e_{g_{i-1}}^{n-1} & \text{if } u_{g_i}^{n-1} > 0 \\ e_{g_{i+1}}^{n-1} - e_{g_i}^{n-1} & \text{otherwise} \end{cases} \tag{7.16}$$

*For particles:*

$$e_{p_i}^n = e_{p_i}^{n-1} - \tfrac{\Delta t}{\Delta x}\left[\bar{u}_p \delta e_p + \tfrac{P_{p_i}^{n-1}}{\bar{u}_p}\left(u_{p_{i+1}}^{n-1} - u_{p_i}^{n-1}\right)\right] + H\Delta t\left(e_{g_i}^{n-1} - e_{p_i}^{n-1}\right) \tag{7.17}$$

where

$$\bar{u}_{p_i}^n = \tfrac{1}{2}\left(u_{p_i}^{n-1} + u_{p_i}^{n-1}\right) \tag{7.18}$$

and

$$\delta e_{p_i}^n = \bar{u}_{p_i}^n \delta \text{energy}_{p_i}^n + \frac{P_{p_i}^{n-1}}{\rho_{p_i}^{n-1}}\left(u_{p_{i+1}}^{n-1} + u_{p_{i+1}}^{n-1}\right) \quad (7.19)$$

$$\delta \text{energy}_{g_i}^n = \begin{cases} e_{p_i}^{n-1} - e_{p_{i-1}}^{n-1} & \text{if } u_{g_i}^{n-1} > 0 \\ e_{p_{i+1}}^{n-1} - e_{p_i}^{n-1} & \text{otherwise} \end{cases} \quad (7.20)$$

**Equation of State**

The ideal gas model was used as equation of state for both phases as follows,

$$P_{g_i}^n = (\gamma_g - 1)\rho_{g_i}^n e_{g_i}^n. \quad (7.21)$$

where $\gamma_g$ is the heat capacity ratio of the gas.

$$P_{p_i}^n = (\gamma_p - 1)\rho_{p_i}^n e_{p_i}^n. \quad (7.22)$$

where $\gamma_p$ is the heat capacity ratio of particles.

This page intentionally left blank.

# Chapter 8

# Two-phase: Verification

*Code verification uses quantitative comparisons of numerical and exact solutions to identify errors in algorithm implementation and algorithmic weaknesses. The primary focus is on the order of convergence of the error as a function of discretization parameters.*

If the friction coefficient (FFF) and the heat transfer coefficient (HHH) are set to zero, we should obtain two independent solutions: one for gas and the other for particles. These results should be exactly the same as those obtained in section 5. The verification was performed by setting the initial velocity to zero for both gas and particles. The gas pressure gradient was set to 0.5 ($P_{g0L}$ = 1 and $P_{g0R}$ = 0.5) and the particles pressure gradient was set to zero ($P_{p0L}$ = 0.5 and $P_{p0R}$ = 0.5). The expected solution, based on the initial conditions, suggests no changes in the initial conditions for particles during the simulation.

In order to calculate the error in each variable, we first determine the corresponding value using the exact solution and then compute the results using the two-phase simulation code. Once we have these two values for all the cells and interfaces, we calculate the error as the sum of the absolute value of the difference between the exact solution and the one-dimensional code, divided by the number of cells/interfaces. This is expressed in Eq. 8.1 for cell quantities and in Eq. 8.2 for edge quantities.

$$error_i = \frac{1}{n} \sum_{i=1}^{n} |y_{exact_i} - y_{sim_i}|, \qquad (8.1)$$

$$error_{i+\frac{1}{2}} = \frac{1}{n} \sum_{i=\frac{1}{2}}^{n+\frac{1}{2}} |y_{exact_{i+\frac{1}{2}}} - y_{sim_{i+\frac{1}{2}}}|, \qquad (8.2)$$

where $y_{exact_i}$ is the result of the exact solution at the point *i*, $y_{sim_i}$ is the simulation code result at the point *i* and *n* is the number of cells considered, similarly for the edges quantities. The procedure was exactly the same that the one used for the single-phase verification in section 5.

As expected, the results match exactly with those obtained in the verification of the single-phase. This indicates that there are no mistakes in the implementation of the equations in the code.

# Chapter 9

# Multi-fidelity Surrogate Model

*High-fidelity (HF) models are accurate but computationally expensive and time-consuming. Low-fidelity (LF) models are inexpensive but less accurate. Multi-fidelity (MF) surrogate models combine both to achieve accuracy at a lower cost. In this work, an MF surrogate is desired to perform an inexpensive uncertainty propagation process.*

## Introduction

The problem modeled is a particle curtain hit by a shock wave. The gas was modeled with an initial discontinuity in pressure (shock tube problem). A discontinuity in density was included where the particle curtain was located. Figure 9.1 shows the initial density distribution of gas and particles.

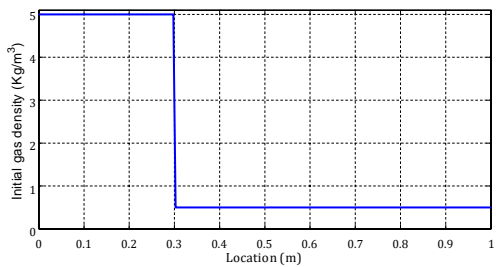

(**a**) Gas initial density distribution.

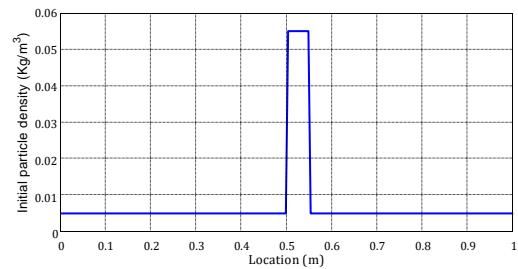

(**b**) Particles initial density distribution.

**Figure 9.1.** Initial conditions in density for a particle curtain hitby a shock wave.

The two-phase simulation code was modified to implement both low-fidelity and high-fidelity simulations. In Section 7, the discretized equations for the gas and particles were described. The modification includes the calculation of the difference and addition of particles and gas velocity. Adding and subtracting Equation 7.7 and Equation 7.8 gives:

$$u_{g_i}^n - u_{p_i}^n$$
$$= \left[ u_{g_i}^{n-1} - \frac{\Delta t}{\Delta x} \left[ u_{g_i}^{n-1} \cdot \delta u_{g_i}^{n-1} + \frac{1}{\bar{\rho}_{g_i}^{n-1}} \left( P_{g_i}^n - P_{g_{i-1}}^n \right) \right] - \frac{F \Delta t}{\bar{\rho}_{g_i}^{n-1}} |u_{g_i}^{n-1} - u_{p_i}^{n-1}|(u_{g_i}^{n-1} - u_{p_i}^{n-1}) \right]$$
$$- \left[ u_{p_i}^{n-1} - \frac{\Delta t}{\Delta x} \left[ u_{p_i}^{n-1} \cdot \delta u_{p_i}^{n-1} + \frac{1}{\bar{\rho}_{p_i}^{n-1}} \left( P_{p_i}^n - P_{p_{i-1}}^n \right) \right] - \frac{F \Delta t}{\bar{\rho}_{p_i}^{n-1}} |u_{p_i}^{n-1} - u_{p_i}^{n-1}|(u_{p_i}^{n-1} - u_{p_i}^{n-1}) \right]$$
(9.1)

$$u_{g_i}^n + u_{p_i}^n$$
$$= \left[ u_{g_i}^{n-1} - \frac{\Delta t}{\Delta x} \left[ u_{g_i}^{n-1} \cdot \delta u_{g_i}^{n-1} + \frac{1}{\bar{\rho}_{g_i}^{n-1}} \left( P_{g_i}^n - P_{g_{i-1}}^n \right) \right] - \frac{F \Delta t}{\bar{\rho}_{g_i}^{n-1}} |u_{g_i}^{n-1} - u_{p_i}^{n-1}|(u_{g_i}^{n-1} - u_{p_i}^{n-1}) \right]$$
$$+ \left[ u_{p_i}^{n-1} - \frac{\Delta t}{\Delta x} \left[ u_{p_i}^{n-1} \cdot \delta u_{p_i}^{n-1} + \frac{1}{\bar{\rho}_{p_i}^{n-1}} \left( P_{p_i}^n - P_{p_{i-1}}^n \right) \right] - \frac{F \Delta t}{\bar{\rho}_{p_i}^{n-1}} |u_{p_i}^{n-1} - u_{p_i}^{n-1}|(u_{p_i}^{n-1} - u_{p_i}^{n-1}) \right].$$
(9.2)

The HF model was built using the HF simulations, i.e., results considering both Equations 9.1 and 9.2. The LF model was built using the LF simulations, i.e., setting Equation 9.2 to zero. This is equivalent to assuming that the difference in velocity of gas and particles is zero during the simulation. To construct the MF model, HF and LF simulations were combined. The detailed process is described in the next section.

## Surrogate models description

The input variables considered were the initial particle curtain thickness ($t_c$) and the initial density jump in the interface ($\rho_{GL}$). Although we considered three possible quantities of interest— the particle curtain maximum density, the particle curtain maximum density location, and the particle curtain standard deviation— we chose just one to perform the surrogate models. The quantity of interest (QoI) considered in this case was the maximum density in the particle curtain due to its higher sensitivity to the input variables. The QoI was calculated at 13 points using the LF simulations, and at the same 13 points using HF simulations. Five of the 13 HF simulation results were selected to build the MF surrogate model.

Most of the time, HF models are computationally prohibitive, and we cannot afford more than a few of them. However, in this case, we want to compare the accuracy of MF models with HF models; therefore, we computed all 13 HF simulations. The sampling points are shown in Figure 9.2.

The LF and HF models were constructed using linear regression with first-order polynomials, including the results of the 13 points obtained using LF and HF simulations, respectively. The MF surrogate model was built by performing a linear regression approximation for the ratio between HF and LF simulations at the 5 selected points. In conclusion, the LF model required 13 LF simulations, while the HF model used all 13 HF simulations, the HF model 13 HF simulations, and the MF model required 13 LF simulation and 5 HF simulations.

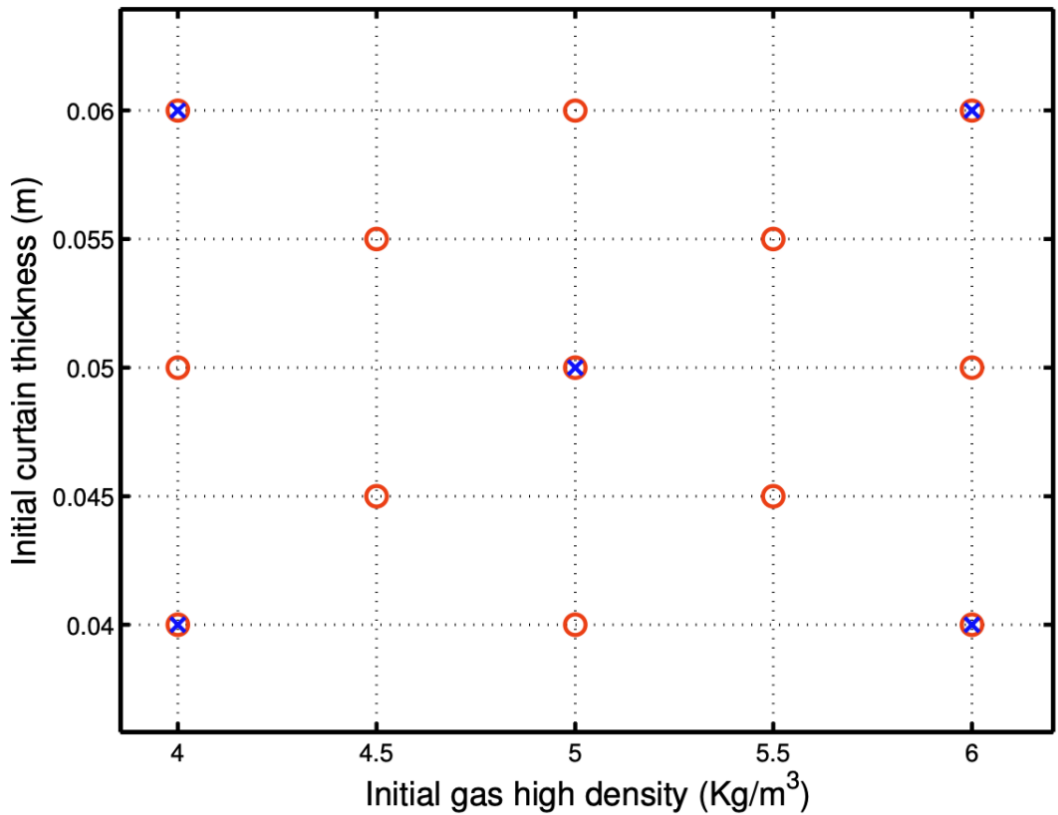

**Figure 9.2.** Red circles LF and HF results were calculated, blue crosses represent the subset of points selected to construct the MF model.

Table 9.1 shows the results of the three surrogate models at each of the 13 selected sampling points. Figure 9.3 graphically compares the performance of the MF surrogate model with the LF and HF models.

In Figure 9.3, it can be observed that the inclusion of the five HF simulation results in the construction of the MF model resulted in an improved model very close to the HF surrogate model, where 13 HF simulation results were used. As a result, we have a better model compared to the LF model, but it is less expensive than the HF model. The code written in MATLAB to build the surrogate models is included in Appendix C.

| $\rho_{GL}(Kg/m^3)$ | $t_c(m)$ | LF | HF | MF | $LF_{diff}(\%)$ | $HF_{diff}(\%)$ |
|---|---|---|---|---|---|---|
| 4.0 | 0.040 | 0.0497 | 0.0431 | 0.0444 | 10.507 | 3.096 |
| 6.0 | 0.040 | 0.0472 | 0.0379 | 0.0357 | 24.283 | 5.871 |
| 5.0 | 0.050 | 0.0551 | 0.0520 | 0.0542 | 1.602 | 4.316 |
| 4.0 | 0.060 | 0.0586 | 0.0738 | 0.0710 | 21.079 | 3.754 |
| 6.0 | 0.060 | 0.0597 | 0.0625 | 0.0641 | 7.303 | 2.569 |
| 4.0 | 0.050 | 0.0552 | 0.0577 | 0.0581 | 5.286 | 0.811 |
| 4.5 | 0.045 | 0.0523 | 0.0474 | 0.0491 | 6.054 | 3.566 |
| 4.5 | 0.055 | 0.0575 | 0.0624 | 0.0631 | 9.738 | 1.115 |
| 5.0 | 0.040 | 0.0484 | 0.0398 | 0.0400 | 17.395 | 0.564 |
| 5.0 | 0.060 | 0.0595 | 0.0673 | 0.0680 | 14.191 | 0.990 |
| 5.5 | 0.045 | 0.0515 | 0.0443 | 0.0449 | 12.942 | 1.228 |
| 5.5 | 0.055 | 0.0576 | 0.0572 | 0.0593 | 2.850 | 3.673 |
| 6.0 | 0.050 | 0.0548 | 0.0489 | 0.0501 | 8.490 | 2.496 |

**Table 9.1.** Results of the LF, HF and MF surrogate models for the selected 13 sampling points.

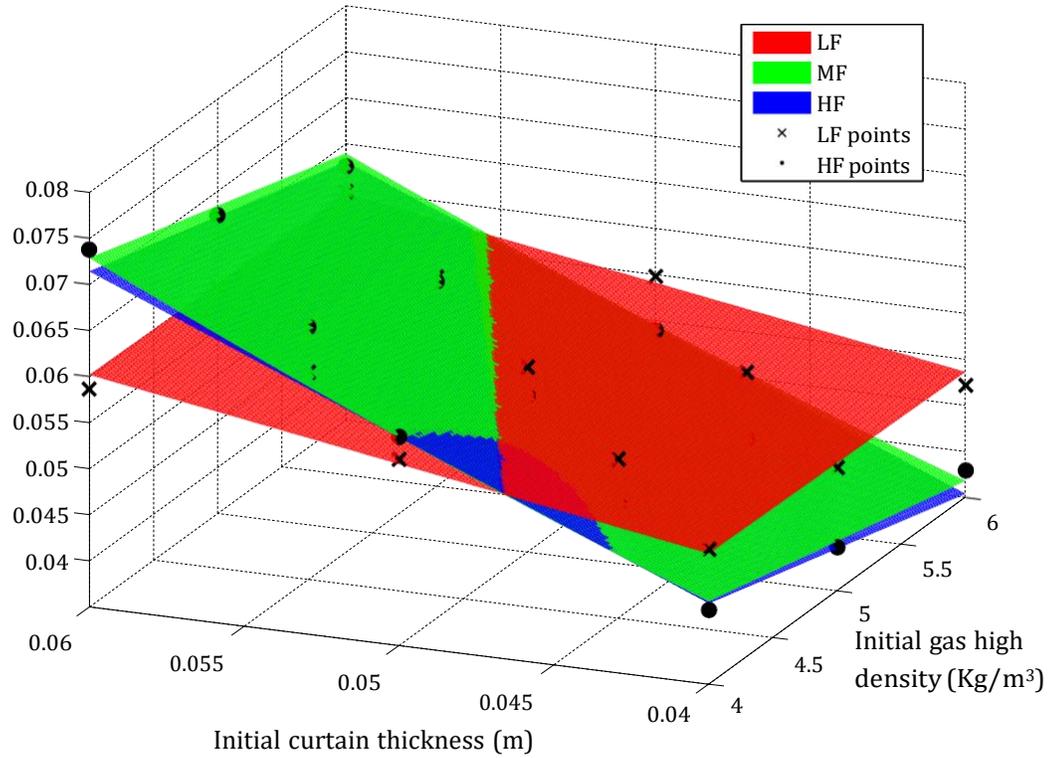

**Figure 9.3.** Red, green and blue plane represents the LF, MF and HF surrogate models respectively. Crosses and dots represent LF and HF simulation results respectively.

This page intentionally left blank.

# Chapter 10

# Uncertainty Propagation

*When variables are the values of experimental measurements, they have uncertainties associated with measurement limitations. In this chapter, measurable uncertainties in outputs due to input uncertainties are identified using the uncertainty propagation process.*

## Inputs Variables and Quantities of Interest

The uncertainty propagation process was applied to the one-dimensional two-phase simulation code. The problem modeled is a particle curtain hit by a shock wave. The gas was modeled with an initial discontinuous pressure gradient (shock tube problem). A density jump was included where the particle curtain is located (see Figure 9.1). Two input variables were selected to perform the propagation: the initial high-density value of the gas, $\rho_{GL}$ and the initial curtain thickness, $t_c$. he quantity of interest (QoI) selected is the particle curtain maximum density. The process was performed using the software Dakota, setting the Latin Hypercube Sampling (LHS) technique with 50 points inside the domain of interest ($4 \leq \rho_{GL} \leq 5$ and $0.04 \leq t_c \leq 0.05$ ).

## Uncertainty Propagation Results

Figure 10.1 shows that the QoI is more sensitive to changes in the initial particle curtain thickness, $t_c$, than to the initial gas high density, $\rho_{GL}$. Uncertainty propagation results are included in Table 10.1. Figure 10.1 graphically represents the results in Table 10.1.

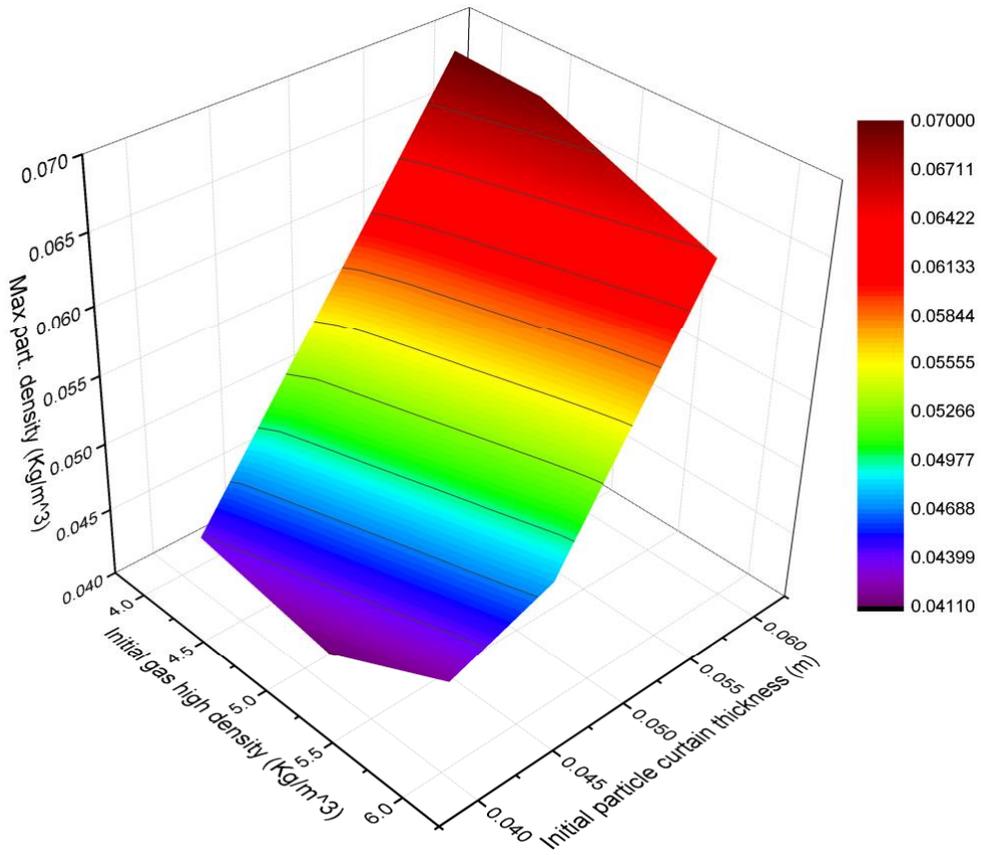

**Figure 10.1.** Variation of the QoI (particle maximum density) by the variation of the selected inputs ($t_c$ and $\rho_{GL}$).

| $\rho_{GL}(Kg/m^3)$ | $t_c(m)$ | $Max.curtain density(Kg/m^3)$ |
|---|---|---|
| 4.77892 | 0.05771 | 0.06579 |
| 5.36759 | 0.04961 | 0.05150 |
| 4.52226 | 0.04895 | 0.05397 |
| 4.36813 | 0.04635 | 0.05100 |
| 4.42620 | 0.05118 | 0.05755 |
| 5.21414 | 0.05565 | 0.06078 |
| 5.50027 | 0.04501 | 0.04485 |
| 4.85095 | 0.05024 | 0.05446 |
| 5.30513 | 0.04689 | 0.04807 |
| 5.88863 | 0.04835 | 0.04773 |
| 5.02582 | 0.04561 | 0.04745 |
| 5.45208 | 0.05461 | 0.05824 |
| 5.53275 | 0.05509 | 0.05860 |
| 5.09785 | 0.04754 | 0.04975 |
| 4.25866 | 0.04672 | 0.05194 |
| 5.36216 | 0.05090 | 0.05331 |
| 5.30430 | 0.04602 | 0.04692 |
| 4.91838 | 0.04451 | 0.04642 |
| 4.81344 | 0.05708 | 0.06467 |
| 5.25450 | 0.05655 | 0.06196 |
| 5.76361 | 0.04998 | 0.05043 |
| 5.24354 | 0.04218 | 0.04222 |
| 4.65031 | 0.04784 | 0.05192 |
| 4.61166 | 0.05925 | 0.06892 |
| 5.01755 | 0.04929 | 0.05247 |
| 5.80575 | 0.05947 | 0.06398 |
| 5.08381 | 0.04089 | 0.04120 |
| 5.15473 | 0.05415 | 0.05883 |
| 4.95644 | 0.05489 | 0.06075 |
| 4.07237 | 0.05837 | 0.06993 |
| 4.53526 | 0.05365 | 0.06070 |
| 5.17284 | 0.04282 | 0.04329 |
| 4.59916 | 0.05349 | 0.06019 |
| 5.56087 | 0.04371 | 0.04297 |
| 4.47551 | 0.05233 | 0.05901 |
| 4.75364 | 0.05668 | 0.06432 |
| 5.40832 | 0.05175 | 0.05433 |
| 4.68397 | 0.05065 | 0.05573 |
| 4.98098 | 0.04408 | 0.04562 |
| 4.90291 | 0.05602 | 0.06267 |
| 4.74428 | 0.05273 | 0.05847 |
| 4.86434 | 0.05853 | 0.06668 |
| 5.64064 | 0.04491 | 0.04419 |
| 5.12485 | 0.04195 | 0.04236 |
| 4.14579 | 0.04027 | 0.04385 |
| 5.69452 | 0.04335 | 0.04201 |
| 4.72273 | 0.05143 | 0.05668 |
| 4.93478 | 0.05313 | 0.05826 |
| 4.31905 | 0.04150 | 0.04478 |
| 5.07049 | 0.04815 | 0.05069 |

**Table 10.1:** Uncertainty propagation of the inputs tc and ρGL into the quantity of interest (particle maximum density).

This page intentionally left blank.

# Chapter 11

# Conclusions

A one-dimensional, single-phase and two-phase simulation code was developed for modeling a shock wave hitting a curtain of particles. Software version control using Git was employed to maintain a record of the different versions of the code. The Eulerian-Eulerian approach was implemented to model both gas and particles.

A verification process was performed to ensure the correct implementation of the equations. Uncertainty propagation was conducted, concluding that the input most affecting the quantity of interest is the initial particle curtain thickness. Since uncertainty propagation requires a large number of simulations, a multi-fidelity surrogate model was built to calculate the selected quantity of interest for the study.

Multi-fidelity surrogate models were constructed to calculate the quantity of interest (maximum density at the particle curtain) using two input variables: initial particle curtain thickness and the initial high-density value.

This page intentionally left blank.

# Appendix A

# Single-phase: Implementation

## A.1 Matlab simulation code

```matlab
%1D 1 Phase     Explicit
clear all;
clc;

global delta_t delta_x firstcell lastcell ncells ntimes nghosts nedges
global firstedge lastedge finaledge finalcell final_time
global xl xr gamma cfl t xc xe

[problem,cell,ncells,nghosts,ntimes,xr,xl,cfl,delta_x, delta_t, total_time, final_time, t]=parameters();

[gamma]=physic();

[xc, xe, nedges, firstcell, lastcell, finalcell, firstedge, lastedge, finaledge]=mesh(ncells);

[rho,P,e,u]=initial(problem, xc, xe);

delta_t=update_t(e(firstcell:lastcell,t) ,u(firstedge:lastedge,t) , total_time, final_time);

while total_time<final_time;
```

```matlab
    t = t + 1;

    %Donor (upwind) density

    delta_rho=upwind_rho(rho,u);

    %conservation of mass equation

    rho(firstcell:lastcell,t)=update_rho(delta_rho,rho);
    rho(:,t)=bc_rho(rho(:,t), problem);

    %Donor (upwind) energy

    delta_e=upwind_e(e,rho,u,P);

    %conservation of energy equation

    e(firstcell:lastcell,t)=update_e(delta_e,e);
    e(:,t)=bc_e(e(:,t), problem);

    %equation of state

    P(:,t)=eos_ideal( rho(:,t), e(:,t));

    %Donor (upwind) velocity

    du2=upwind_u(u,P,rho);

    %conservation of momentum equation

    u(firstedge:lastedge,t)=update_u(du2,u);
    u(:,t)=bc_u(u(:,t), problem);

    % timestep
    delta_t=update_t(e(firstcell:lastcell,t) ,u(firstedge:lastedge,t), total_time, final_time);
    total_time=total_time+delta_t;

end
```

# Appendix B

# Two Phases: Implementation

## B.1 Two Phases

A second phase was implemented, including momentum exchange due to friction and energy exchange due to heat transfer between phases. The structure of the MATLAB code written is shown in the following lines.

## B.2 Matlab simulation code

```matlab
%1D 2 Phase    Explicit

[problem, cell, ncells, nghosts, ntimes, xr, xl, cfl, delta_x, delta_t,
    total_time, final_time, t, PCT, rhogl, rhopl, model]= parameters();

[gamma, H, F]= physic();

[xc, xe, nedges, firstcell, lastcell, finalcell, firstedge,
    lastedge, finaledge]= mesh(ncells);

[rho_gas, P_gas, e_gas, u_gas]= initial(problem, xc, xe, 1);

[rho_part, P_part, e_part, u_part]= initial(problem, xc, xe, 2);

[u_gp_diff, u_gp_add, e_gp_diff, e_gp_add]= diff_add(u_gas, u_part
    , e_gas, e_part);

```

```matlab
15    delta_t=update_t(e_gas(firstcell:lastcell,t), u_gas(firstedge+1:lastedge+1,t), total_time, finaltime);
16
17 while total_time<final_time
18
19     t=t+1;
20
21     %Donor (upwind) density
22
23     delta_rho_gas=upwind_rho(rho_gas, u_gas);
24     delta_rho_part=upwind_rho(rho_part, u_part);
25
26
27     %conservation of mass equation
28
29     rho_gas(firstcell:lastcell,t)=update_rho(delta_rho_gas, rho_gas);
30     rho_part(firstcell:lastcell,t)=update_rho(delta_rho_part, rho_part);
31     rho_gas(:,t)=bc_rho(rho_gas(:,t), problem);
32     rho_part(:,t)=bc_rho(rho_part(:,t), problem);
33
34
35     %Donor (upwind) energy
36
37     delta_e_gas=upwind_e(e_gas, rho_gas, u_gas, P_gas);
38     delta_e_part=upwind_e(e_part, rho_part, u_part, P_part);
39
40
41     %conservation of energy equation
42
43     [e_gp_diff, e_gp_add]=update_e_diff_add(delta_e_gas, delta_e_part, e_gas, e_part, rho_gas, rho_part, e_gp_diff, e_gp_add);
44     [e_gas, e_part]=calculate_e(e_gp_diff, e_gp_add, e_gas, e_part);
45     e_gas(:,t)=bc_e(e_gas(:,t), problem);
46     e_part(:,t)=bc_e(e_part(:,t), problem);
47
48     %equation of state
49
50     P_gas(:,t)=eos_ideal(rho_gas(:,t), e_gas(:,t));
51     P_part(:,t)=eos_ideal(rho_part(:,t), e_part(:,t));
52
53
54     %Donor (upwind) velocity
```

```matlab
        du2_gas=upwind_u(u_gas,P_gas,rho_gas);
        du2_part=upwind_u(u_part,P_part,rho_part);

    %conservation of momentum equation

        [u_gp_diff,u_gp_add]=update_u_diff_add(du2_gas,du2_part,u_gas,
            u_part,rho_gas,rho_part,u_gp_diff,u_gp_add,model,srgtSRGT,B
            );
        [u_gas,u_part]=calculate_u(u_gp_diff,u_gp_add,u_gas,u_part);
        u_gas(:,t)=bc_u(u_gas(:,t), problem);
        u_part(:,t)=bc_u(u_part(:,t), problem);

    %time step calculation

        delta_t1=update_t(e_gas(firstcell:lastcell,t) ,u_gas(firstedge
            1:lastedge+1,t), total_time, finaltime);
        delta_t2=update_t(e_part(firstcell:lastcell,t) ,u_part(
            firstedge 1:lastedge+1,t), total_time, finaltime);

        delta_t=min(delta_t1, delta_t2);

        total_time=total_time+delta_t;

end
```

This page intentionally left blank.

# Appendix C

# Surrogate Models: Implementation

*Surrogate models are used to reduce the cost of simulations when a high number of simulations are needed. The surrogate models implemented were coded in MATLAB and are included in this section. Low-fidelity, high-fidelity, and multi-fidelity surrogate models of the 1D-2Phase code were built, and the MATLAB code is included below. Detailed information about the surrogate models can be found in Chapter 9.*

```matlab
%Inputs rho_gl:density before the shock, tc= curtain thickness, output >
%location of the particle maximun density

[B_LF, B_RMF, B_HF]=models();

n=100;
rho_gl=linspace(4,6,n)';
tc=linspace(0.04,0.06,n)';
for i=1:n
    for j=1:n
        X1mat(i,j) = rho_gl(i);
        X2mat(i,j) = tc(j);
        Y_LF(i,j) = [1, rho_gl(i), tc(j)] B_LF;
        Y_HF(i,j) = [1, rho_gl(i), tc(j)] B_HF;
        RY(i,j) = [1, rho_gl(i), tc(j)] B_RMF;
    end
```

```matlab
18  end
19
20  X=[ones(n,1)  rho_gl  tc];
21
22  Y_MF=RY.*Y_LF;
23
24  % Sampling points
25  X1=[4      0.04
26  6        0.04
27  5        0.05
28  4        0.06
29  6        0.06
30  4        0.05
31  4.5      0.045
32  4.5      0.055
33  5        0.04
34  5        0.06
35  5.5      0.045
36  5.5      0.055
37  6        0.05];
38
39
40  Y_HF1=[0.04310066
41  0.0379363
42  0.05195248
43  0.07375866
44  0.06249921
45  0.05765025
46  0.04742803
47  0.06239245
48  0.03975291
49  0.06732769
50  0.04433359
51  0.0571908
52  0.04890361];
53
54  Y_LF1=[0.0496518
55  0.04716079
56  0.05507725
57  0.05863083
58  0.05974177
59  0.05520001
60  0.0522849
61  0.05748953
62  0.04839559
```

```matlab
63    0.05954426
64    0.0515495
65    0.05764807
66    0.05477452];
67
68
69  figure;
70  surf( X1mat, X2mat, Y_LF );
71  hold on;
72  surf( X1mat, X2mat, Y_MF );
73  hold on;
74  surf( X1mat, X2mat, Y_HF );
75  hold on;
76  scatter3(X1(:,1), X1(:,2), Y_LF1,'+m');
77  hold on;
78  scatter3(X1(:,1), X1(:,2), Y_HF1,'+k');
79
80  function [B_LF, B_RMF, B_HF]=models()
81
82  % Common points (used in MF)
83  X1=[4      0.04
84      6      0.04
85      5      0.05
86      4      0.06
87      6      0.06];
88
89  X1_extended =[ones(length(X1),1) X1(:,1)  X1(:,2)];
90
91  % Sampling points
92  X2=[
93      4      0.04
94      6      0.04
95      5      0.05
96      4      0.06
97      6      0.06
98      4      0.05
99      4.5    0.045
100     4.5    0.055
101     5      0.04
102     5      0.06
103     5.5    0.045
104     5.5    0.055
105     6      0.05];
106
107 X2_extended =[ones(length(X2),1) X2(:,1)  X2(:,2)];
```

```matlab
%HF results
Y_HF1=[0.04310066
0.0379363
0.05195248
0.07375866
0.06249921];

Y_HF2=[0.04310066
0.0379363
0.05195248
0.07375866
0.06249921
0.05765025
0.04742803
0.06239245
0.03975291
0.06732769
0.04433359
0.0571908
0.04890361];

%LF results
Y_LF1=[0.0496518
0.04716079
0.05507725
0.05863083
0.05974177];

Y_LF2=[0.0496518
0.04716079
0.05507725
0.05863083
0.05974177
0.05520001
0.0522849
0.05748953
0.04839559
0.05954426
0.0515495
0.05764807
0.05477452];

% Surrogate for LF
```

```matlab
153  B_LF=regress(Y_LF2, X2_extended);
154
155  % Ratio of HF to LF on X1
156
157  RY1=Y_HF1./Y_LF1;
158
159  % Surrogate for the ratio
160
161  B_RMF=regress(RY1, X1_extended);
162
163  %HF surrogate for comparison
164
165  B_HF=regress(Y_HF2, X2_extended);
```

This page intentionally left blank.

This page intentionally left blank.

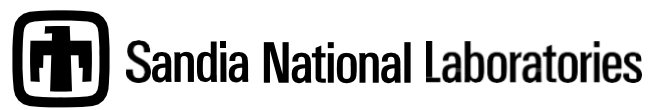